\documentclass{aa} 
\usepackage{graphicx}
\usepackage{txfonts}

\begin{document}

   \title{Tracking the 3D evolution of a halo coronal mass ejection using the revised cone model}

   \author{Q. M. Zhang\inst{1}}

   \institute{Key Laboratory of Dark Matter and Space Astronomy, Purple Mountain Observatory, CAS, Nanjing 210023, PR China \\
                  \email{zhangqm@pmo.ac.cn}}

   \date{Received; accepted}
   \titlerunning{Tracking the 3D evolution of a halo coronal mass ejection}
   \authorrunning{Q. M. Zhang}
 
  \abstract 
   {}
   {This paper aims to track the three-dimensional (3D) evolution of a full halo coronal mass ejection (CME) on 2011 June 21.}
   {The CME results from a non-radial eruption of a filament-carrying flux rope in NOAA active region 11236.
   The eruption is observed in extreme-ultraviolet (EUV) wavelengths by the Extreme-Ultraviolet Imager (EUVI) on board the 
   ahead and behind Solar TErrestrial RElations Observatory (STEREO) spacecrafts and the Atmospheric Imaging Assembly (AIA) on board the Solar Dynamics Observatory (SDO).
   The CME is observed by the COR1 coronagraph on board STEREO and the C2 coronagraph on board the Large Angle Spectroscopic Coronagraph (LASCO).
   The revised cone model is slightly modified, with the top of the cone becoming a sphere, which is internally tangent to the legs.
   Using the multi-point observations, the cone model is applied to derive the morphological and kinematic properties of the CME.}
   {The cone shape fits nicely with the CME observed by EUVI and COR1 on board STEREO twin spacecraft and LASCO/C2 coronagraph.
    The cone angle increases sharply from 54$^{\circ}$ to 130$^{\circ}$ in the initial phase, indicating a rapid expansion.
    A relation between the cone angle and heliocentric distance of CME leading front is derived, $\omega=130\degr-480d^{-5}$, where $d$ is in unit of $R_{\odot}$.
     The inclination angle decreases gradually from $\sim$51$^{\circ}$ to $\sim$18$^{\circ}$, suggesting a trend of radial propagation.
     The heliocentric distance  increases gradually in the initial phase and quickly in the later phase up to $\sim$11\,$R_{\odot}$.
     The true speed of CME reaches $\sim$1140 km s$^{-1}$, which is $\sim$1.6 times higher than the apparent speed in the LASCO/C2 field of view.}
   {The revised model is promising in tracking the complete evolution of CMEs.}
   
   \keywords{Sun: coronal mass ejections (CMEs) -- Sun: flares -- Sun: filaments, prominences}

   \maketitle

\section{Introduction} \label{intro}
Solar flares \citep{fle11} and coronal mass ejections \citep[CMEs;][]{chen11,pat20} are the most powerful activities in the solar atmosphere where 
a huge mount of magnetic free energy is impulsively released \citep{asch17}. A temporal relationship is found between CMEs and the associated flares \citep{zj01}.
The initial phase, impulsive acceleration phase, and propagation phase of CMEs are closely related to the pre-flare phase, rise phase, and decay phase of flares, respectively.
It is widely accepted that magnetic flux ropes play an essential role in driving CMEs and flares \citep{chen03,au10,zj12,cx13,jan13,vour13,sa20,zqm22}.

The typical shape of a CME is the well-known three-part structure: a bright core within a dark cavity surrounded by a bright leading front \citep[e.g.,][]{ill85,ill86,cre04,ver18,zqm18,dai21}.
The frontside halo CMEs, originating near the solar disk center and propagating toward the Earth \citep{how82,go07,by10,zqm10,zqm17,lu17,yan21}, may have severe impact on the solar-terrestrial 
environment \citep[see][and references therein]{tem21a}. A positive correlation is found between the radial speed and the lateral expansion speed of halo CMEs \citep{sch05}.
The three-dimensional (3D) morphology and kinematics of halo CMEs are critical for an accurate estimation of the arrival time \citep{shen21}.
Under the assumption that the velocity and angular width keep constant, 
several versions of cone models were proposed to quickly determine the geometry and kinematics of halo CMEs \citep[e.g.,][]{zhao02,mich03,xie04,xue05,mich06}. 
Later on, a more sophisticated graduated cylindrical shell \citep[GCS;][]{the06,the09,the11} model was developed to carry out forward modeling of flux-rope-like CMEs.
The GCS model is characterized by six geometric parameters: the Carrington longitude ($\phi$) and latitude ($\theta$) of the source region, the tilt angle ($\gamma$) of the flux rope,
the angular width (2$\alpha$), aspect ratio ($\kappa$), and height ($h$) of the legs, respectively.
The model has been successfully applied to tracking the morphology evolution of CMEs using multi-instrument observations \citep[e.g.,][]{pat10a,tem12,tem21b,cx14,col15,cab16,liu18,gou20,ok21}.
To determine the 3D structure of two fronts of a CME, \citet{kw14} put forward a compound model: an ellipsoid model representing the bubble-shaped shock structure \citep{rou16}
and a GCS model representing the flux rope-shaped structure. \citet{is16} created a novel 3D analytic model of CMEs, 
which is able to reproduce the global shape of a CME with all major deformations. 
Using the time-dependent, self-similar Gibson-Low model \citep{gib98}, \citet{dai22} reconstructed the CME on 2011 March 7 and derived the size, shape, velocity, and magnetic field strength.

To investigate the 3D evolution of CMEs as a result of non-radial filament eruptions, \citet{zqm21} proposed a revised cone model (see Fig. 1).
The apex of the axisymmetric cone is located at the source region of an eruption instead of the Sun center in the traditional cone models. 
The model has four geometric parameters: the length ($r$) and angular width ($\omega$) of the cone, an inclination angle ($\theta_1$) from the local vertical, 
and a deviation angle ($\phi_1$) from the local meridian plane. 
The model was applied to two partial halo CMEs originating from the western limb on 2011 August 11 and 2012 December 7.
The values of $\theta_1$ reach up to 70$^{\circ}$ and 60$^{\circ}$. Both CMEs are off-pointed by 30$^{\circ}$ with respect to the plane of the sky.
However, the preliminary application is restricted to the very early evolution of CMEs. 

On 2011 June 21, a filament-carrying flux rope erupted non-radially from NOAA active region (AR) 11236, giving rise to a C7.7 class long-duration flare and a full halo CME (AW$=$360$^\circ$).
\citet{zhou17} tracked the 3D evolution of the eruption by using simultaneous observations from the Atmospheric Imaging Assembly \citep[AIA;][]{lem12} 
on board the Solar Dynamics Observatory (SDO) and the Extreme-Ultraviolet Imager \citep[EUVI;][]{wu04}
of the Sun-Earth Connection Coronal and Heliospheric Investigation \citep[SECCHI;][]{how08} instrument
on board the ahead and behind Solar TErrestrial RElations Observatory \citep[STEREO;][]{kai08}.
\citet{guo19} obtained a magnetic flux rope before eruption, with the locations of magnetic dips being cospatial with part of the filament/prominence material.
Moreover, \citet{guo21} performed a data-constrained magnetohydrodynamic (MHD) simulation of the flux rope eruption, which is consistent with the multi-perspective observations.
In this paper, the revised cone model is slightly modified and applied to tracking the 3D evolution of the halo CME on 2011 June 21. The paper is organized as follows.
The method and data analysis are described in Sect.~\ref{data}. The results are presented in Sect.~\ref{result} and compared with previous works in Sect.~\ref{dis}.
A brief summary is given in Sect.~\ref{sum}.

\section{Method and data analysis} \label{data}
First, let us recall the model \citep[][see Fig. 1]{zqm21}.
The Carrington longitude and latitude of the source region of filament eruption are denoted by $\phi_2$ and $\beta_2$ ($\theta_2=90^\circ-\beta_2$), respectively. 
Hence, the transform between the heliocentric coordinate system (HCS; $X_h$, $Y_h$, $Z_h$) and local coordinate system (LCS; $X_l$, $Y_l$, $Z_l$) 
is as follows \footnote{There is an unintended error in Eqn. (1) of \citet{zqm21}.}:

\begin{equation} \label{eqn-1}
\left(
\begin{array}{c}
x_h  \\
y_h  \\
z_h \\
\end{array}
\right)
=M_2
\left(
\begin{array}{c}
x_l  \\
y_l  \\
z_l \\
\end{array}
\right)
+
\left(
\begin{array}{c}
R_{\odot}\sin{\theta_2}\cos{\phi_2} \\
R_{\odot}\sin{\theta_2}\sin{\phi_2} \\
R_{\odot}\cos{\theta_2} \\
\end{array}
\right),
\end{equation}
where
\begin{equation} \label{eqn-2}
M_2=
\left(
\begin{array}{ccc}
\cos{\theta_2}\cos{\phi_2} & -\sin{\phi_2} & \cos{\phi_2}\sin{\theta_2}  \\
\cos{\theta_2}\sin{\phi_2}  &  \cos{\phi_2} & \sin{\phi_2}\sin{\theta_2} \\
-\sin{\theta_2}                    &         0          &  \cos{\theta_2} \\
\end{array}
\right).
\end{equation}

The transform between LCS and cone coordinate system (CCS; $X_c$, $Y_c$, $Z_c$) is performed by a matrix ($M_1$):

\begin{equation} \label{eqn-3}
\left(
\begin{array}{c}
x_l  \\
y_l  \\
z_l \\
\end{array}
\right)
=M_1
\left(
\begin{array}{c}
x_c  \\
y_c  \\
z_c \\
\end{array}
\right),
\end{equation}
where
\begin{equation} \label{eqn-4}
M_1=
\left(
\begin{array}{ccc}
\cos{\theta_1}\cos{\phi_1} & -\sin{\phi_1} & \cos{\phi_1}\sin{\theta_1} \\
\cos{\theta_1}\sin{\phi_1}  &  \cos{\phi_1} & \sin{\phi_1}\sin{\theta_1} \\
-\sin{\theta_1}                    &         0          &  \cos{\theta_1} \\
\end{array}
\right).
\end{equation}

\begin{figure}
\centering
\includegraphics[width=9cm]{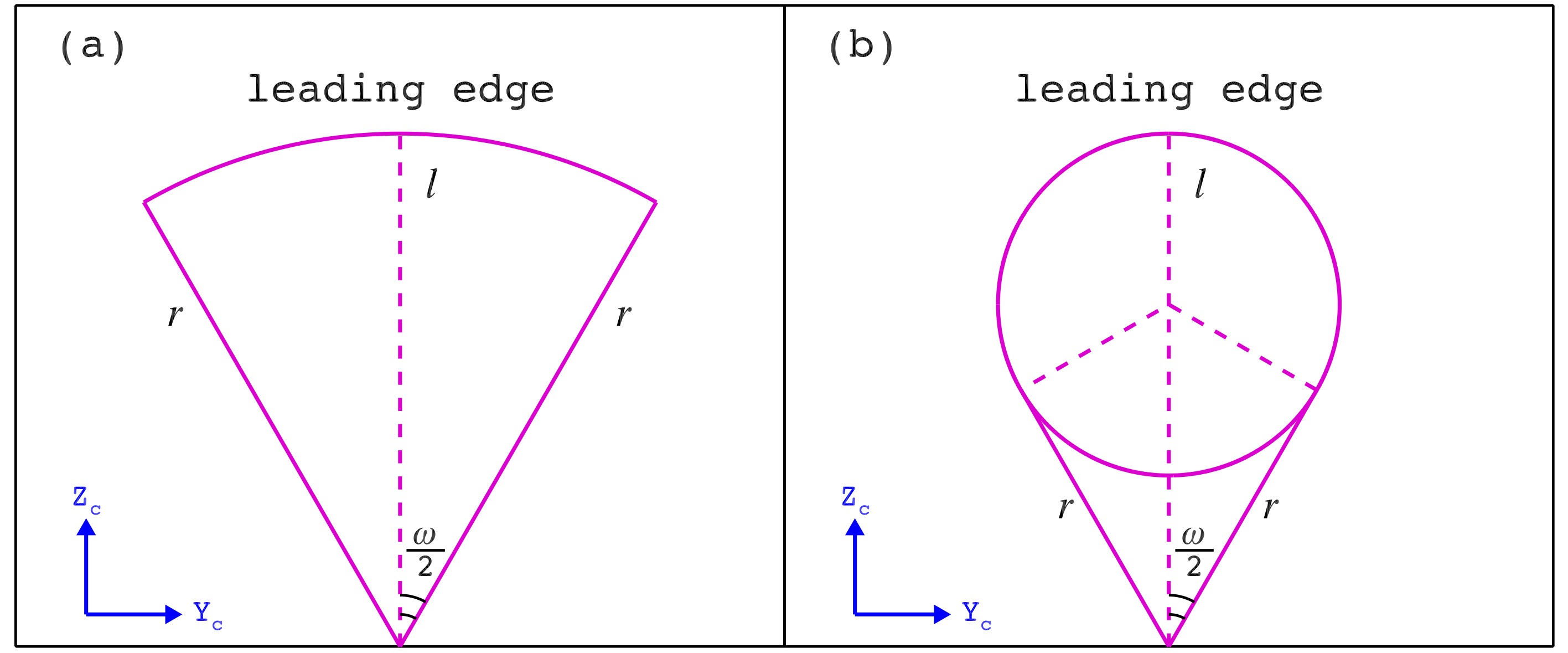}
\caption{(a) Cross section of the cone in the $Y_c$-$Z_c$ plane \citep{zqm21}. (b) Cross section of the modified cone in the $Y_c$-$Z_c$ plane in this paper.} 
\label{fig1}
\end{figure}

In the revised model, the base of cone is a sphere section \citep[see Model A in][]{sch05}. The cone has a length of $r$ and angular width of $\omega$. 
Therefore, the cross section of the cone is fan-shaped in the $Y_c$-$Z_c$ plane, and the leading edge has a total length of $l=r$ (Fig.~\ref{fig1}(a)). 
In view of the flux rope nature of CMEs in many cases, there is a need to modify the shape of the cone. 
The top of the cone becomes a sphere, which is internally tangent to the legs \citep[see Model C in][]{sch05}.
Fig.~\ref{fig1}(b) shows the cross section of the modified cone in the $Y_c$-$Z_c$ plane, which is similar to the previous models \citep[e.g.,][]{the06}.
It is obvious that the angular width ($\omega$) is the same, while the total length of the leading edge is:
\begin{equation} \label{eqn-5}
l=r(\tan{\frac{\omega}{2}}+(\cos{\frac{\omega}{2}})^{-1}).
\end{equation}

The aspect ratio ($\kappa$) of a CME bubble is defined as the ratio of center height to radius \citep{pat10b,ver18}. In Fig.~\ref{fig1}(b), $\kappa=(\sin{\frac{\omega}{2}})^{-1}$.
To determine the geometric parameters of the modified model, observations from multiple viewpoints are required. 
For the twin satellites of STEREO, the separation angles from the Sun-Earth connection are 
$\phi_{0A}\approx95^{\circ}$ for the ahead (STA) and $\phi_{0B}\approx-92^{\circ}$ for the behind (STB) spacecraft on 2011 June 21 \citep{zhou17}.

The transform between the STA coordinate system ($X_{ea}$, $Y_{ea}$, $Z_{ea}$) and HCS is performed by a matrix ($M_{0A}$):
\begin{equation} \label{eqn-6}
\left(
\begin{array}{c}
x_{ea}  \\
y_{ea}  \\
z_{ea} \\
\end{array}
\right)
=M_{0A}
\left(
\begin{array}{c}
x_h  \\
y_h  \\
z_h \\
\end{array}
\right),
\end{equation}
where
\begin{equation} \label{eqn-7}
M_{0A}=
\left(
\begin{array}{ccc}
\cos{\phi_{0A}} & \sin{\phi_{0A}} & 0 \\
-\sin{\phi_{0A}} & \cos{\phi_{0A}} & 0 \\
0                  &     0              & 1 \\
\end{array}
\right).
\end{equation}

Likewise, the transform between the STB coordinate system ($X_{eb}$, $Y_{eb}$, $Z_{eb}$) and HCS is performed by a matrix ($M_{0B}$):
\begin{equation} \label{eqn-8}
\left(
\begin{array}{c}
x_{eb}  \\
y_{eb}  \\
z_{eb} \\
\end{array}
\right)
=M_{0B}
\left(
\begin{array}{c}
x_h  \\
y_h  \\
z_h \\
\end{array}
\right),
\end{equation}
where
\begin{equation} \label{eqn-9}
M_{0B}=
\left(
\begin{array}{ccc}
\cos{\phi_{0B}} & \sin{\phi_{0B}} & 0 \\
-\sin{\phi_{0B}} & \cos{\phi_{0B}} & 0 \\
0                  &     0              & 1 \\
\end{array}
\right).
\end{equation}

The eruptive prominence and associated flare were observed by STA and STB in EUVI 195 {\AA} and 304 {\AA} images with a cadence of 5 minutes.
The $\mathsf{S}$-shaped, filament-carrying flux rope was observed by SDO/AIA in 94 {\AA} ($T\approx6.3$ MK) and 211 {\AA} ($T\approx2$ MK).
The associated CME was observed by COR1 coronagraph on board STEREO with a cadence of 5 minutes 
and by C2 coronagraph of the Large Angle Spectroscopic Coronagraph \citep[LASCO;][]{bru95} 
instrument on board SOHO \footnote{https://cdaw.gsfc.nasa.gov/CME\_list/} with a cadence of $\geq$10 minutes. 
The field of views (FOVs) of SECCHI/COR1 and LASCO/C2 are 1.5$-$4\,$R_{\odot}$ and 2$-$6\,$R_{\odot}$, respectively.
The soft X-ray (SXR) fluxes of the long-duration flare in 0.5$-$4 {\AA} and 1$-$8 {\AA} were observed by the GOES spacecraft.
The photospheric line-of-sight (LOS) magnetograms before eruption were observed by the Helioseismic and Magnetic Imager \citep[HMI;][]{sch12} on board SDO.
The global 3D magnetic configuration before eruption was derived from the potential-field source surface \citep[PFSS;][]{sch03} modeling.

\section{Results} \label{result}
\begin{figure}
   \centering
   \includegraphics[width=8.5cm]{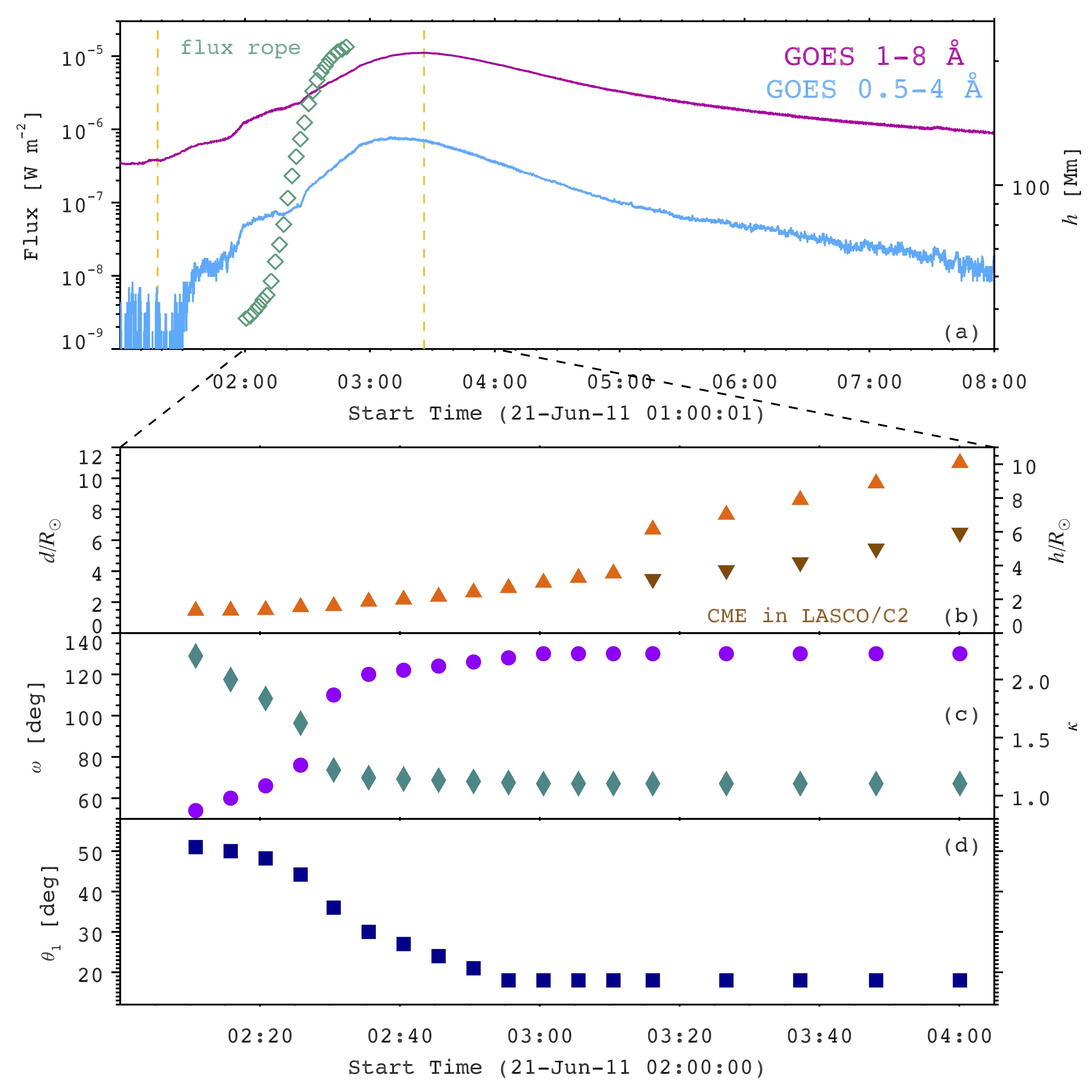}
      \caption{(a) SXR light curves of the flare in 1$-$8 {\AA} (magenta line) and 0.5$-$4 {\AA} (blue line).
      The yellow dashed lines represent the start ($\sim$01:18 UT) and peak ($\sim$03:26 UT) times of the flare.
      The green diamonds denote the heights of magnetic flux rope in the MHD simulation \citep{guo21}.
      (b-d) Time evolutions of the real distance ($d$) of the CME leading edge, angular width ($\omega$), aspect ratio ($\kappa$), and inclination angle ($\theta_1$) of the cone.
      In panel (b), the heights of CME in LASCO/C2 FOV are drawn with chocolate triangles.}
      \label{fig2}
   \end{figure}

In Fig.~\ref{fig2}, the top panel shows the SXR light curves of the flare. The SXR flux in 1$-$8 {\AA} increases from $\sim$01:18 UT to the peak value at $\sim$03:26 UT before decreasing gradually.
During the impulsive phase, the constructed flux rope (green diamonds) rises quickly from $\sim$02:00 UT until $\sim$02:48 UT \citep{guo21}.

\begin{figure}
   \centering
   \includegraphics[width=8.5cm]{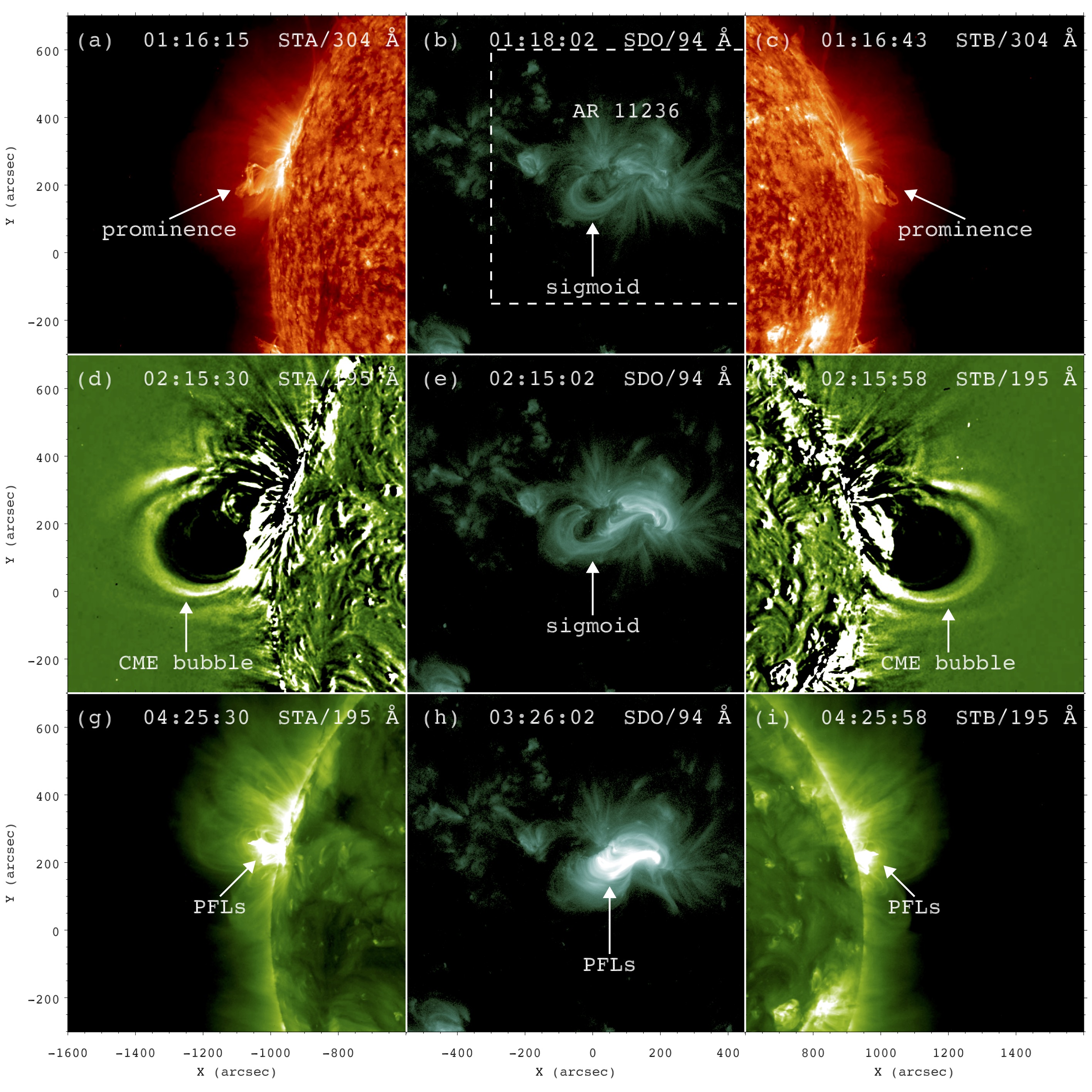}
      \caption{The prominence observed by STA (a) and STB (c) in 304 {\AA} before eruption.
      The CME bubble observed by STA (d) and STB (f) in 195 {\AA} base-difference images during eruption.
      Post-flare loops (PFLs) observed by STA (g) and STB (i) in 195 {\AA} after eruption.
      The AIA 94 {\AA} images feature the sigmoid before (b) and during (e) eruption, and the PFLs (h).
      The white dashed box in panel (b) signifies the FOV of Fig.~\ref{fig4}(a).}
      \label{fig3}
   \end{figure}

In Fig.~\ref{fig3}, the top panels show the bright prominence observed by STEREO/EUVI in 304 {\AA} and the sigmoid observed by SDO/AIA in 94 {\AA} before eruption \citep{zhou17}.
The middle panels show the CME bubble observed by EUVI in 195 {\AA} (base-difference) and the sigmoid in 94 {\AA} during eruption.
The bottom panels show the hot and bright post-flare loops (PFLs) after eruption.

\begin{figure}
\centering{
            \includegraphics[width=0.22\textwidth,clip=]{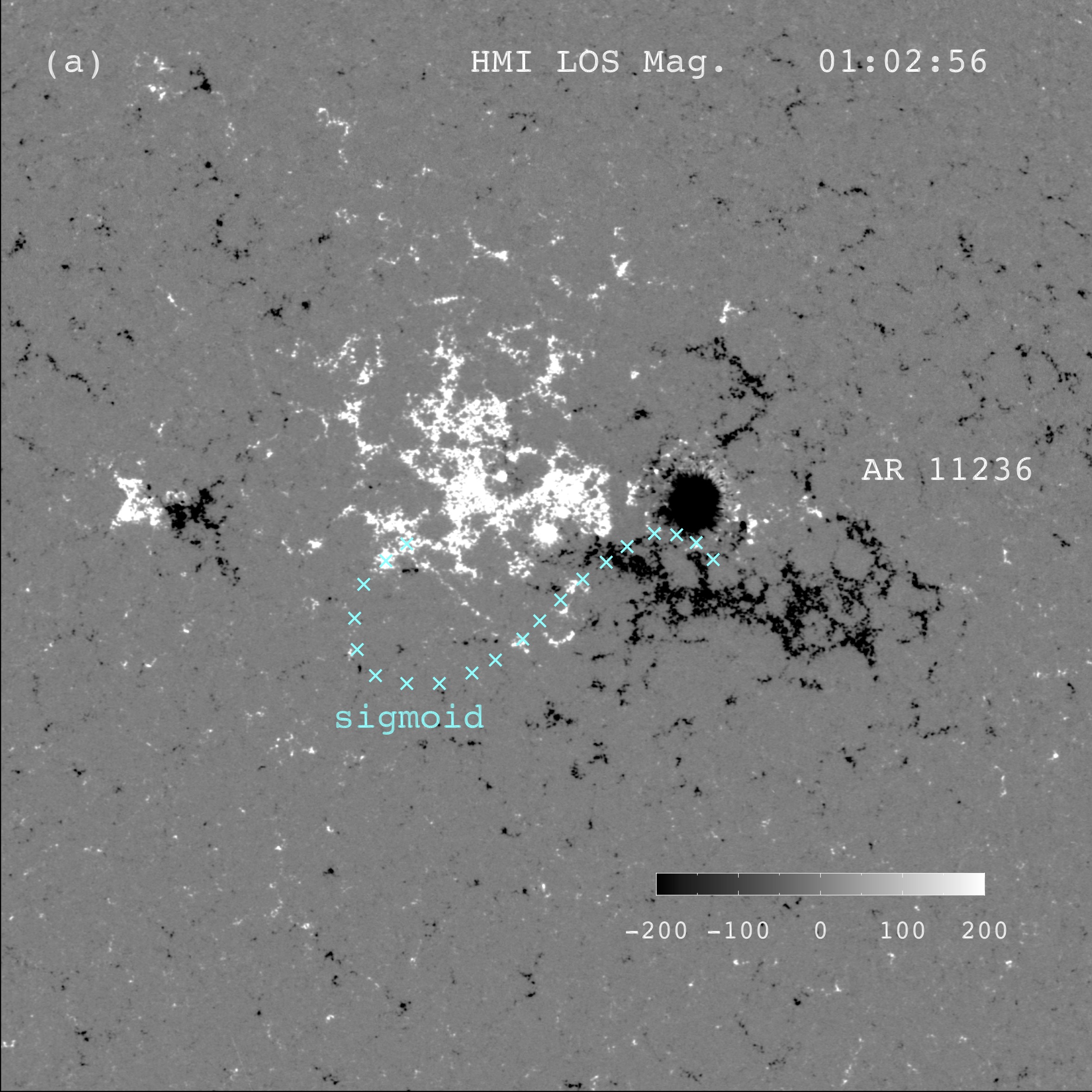}
            \includegraphics[width=0.22\textwidth,clip=]{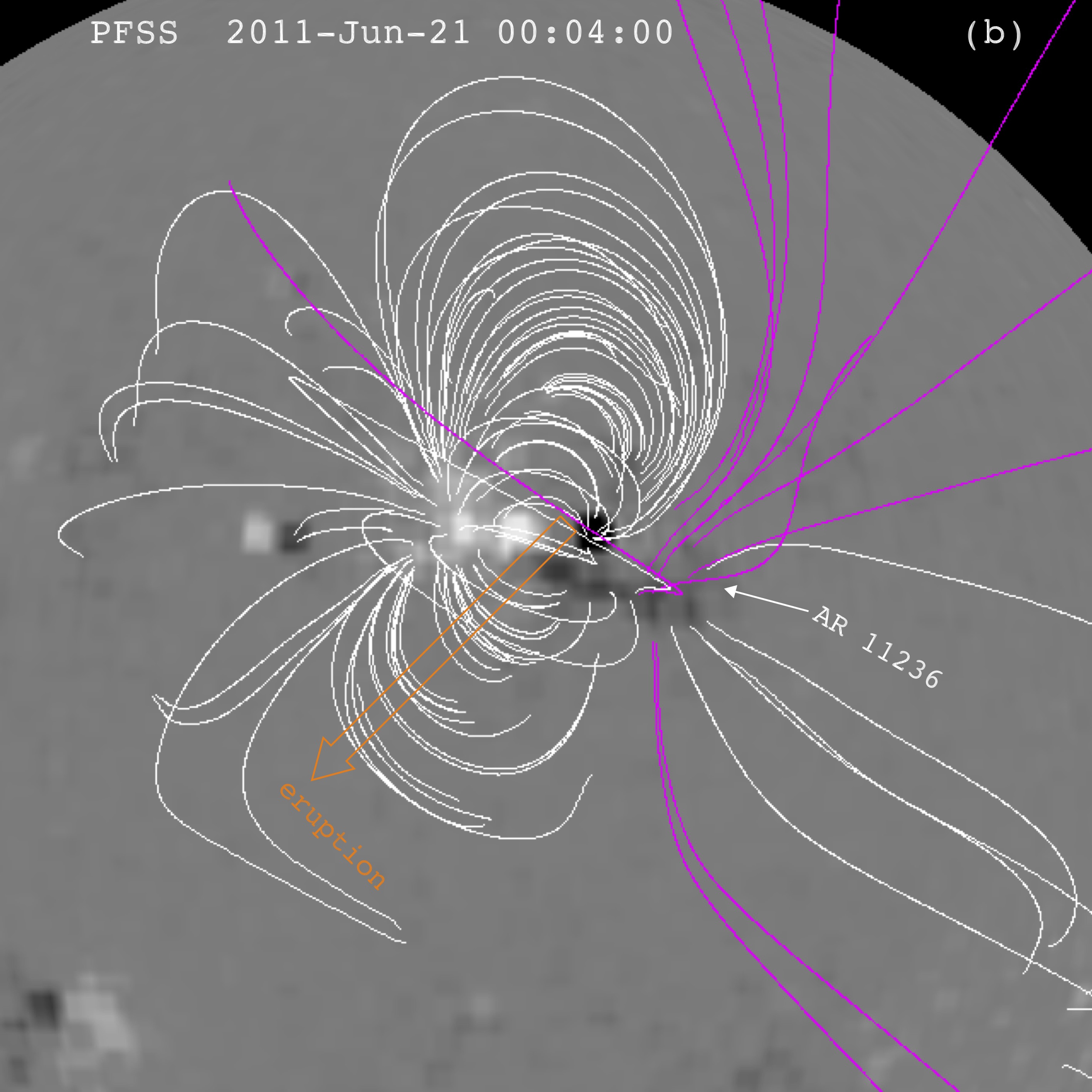}
           }
\caption{(a) The LOS magnetogram of AR 11236 at 01:02:56 UT. The cyan crosses represent the locations of sigmoid in 94 {\AA}.
(b) The large-scale 3D magnetic configuration around the AR at 00:04 UT derived from PFSS modeling.
The white and magenta lines represent the closed and open field lines. The orange arrow indicates the projected direction of filament eruption.}
\label{fig4}
\end{figure}

In Fig.~\ref{fig4}, the left panel shows the LOS magnetogram of AR 11236 at 01:02:56 UT, with the locations of sigmoid being superposed with cyan crosses.
The sigmoid is roughly along the polarity inversion line.
The right panel shows the large-scale 3D magnetic configuration around the AR at 00:04 UT using PFSS modeling. The orange arrow indicates the projected direction of filament eruption.
It is revealed that the AR is adjacent to open field lines (magenta) that may deflect the eruption in longitude \citep{cre06,kil09,pan13,yang18}.
  
\begin{table}
\centering
\caption{Parameters of the cones at different times.
Note that $\phi_2=8.5^{\circ}$, $\beta_2=16.9^{\circ}$, and $\phi_1=-45^{\circ}$ keep constant during the evolution.} 
\label{tab-1}
\begin{tabular}{ccccccc}
\hline\hline
Time & $r$ & $\omega$ & $\theta_1$ & $l$ & $d$ & $\kappa$ \\
(UT)  &  ($\arcsec$) & ($^{\circ}$) & ($^{\circ}$) & ($\arcsec$) & ($R_\odot$) &  \\
\hline
02:10 & 340$\pm$5 & 54$\pm$1 & 51.0$\pm$0.2 &  555 & 1.44 & 2.20 \\
02:15 & 325$\pm$4 & 60$\pm$1 & 50.0$\pm$0.2 &  563 & 1.46 & 2.00 \\
02:20 & 323$\pm$4 & 66$\pm$1 & 48.2$\pm$0.2 &  595 & 1.50 & 1.84 \\
02:25 & 375$\pm$6 & 76$\pm$1 & 44.2$\pm$0.2 &  769 & 1.68 & 1.62 \\
\hline
02:30 & 250$\pm$2 & 110$\pm$2 & 36.0$\pm$0.5 &    793 & 1.75 & 1.22 \\
02:35 & 280$\pm$3 & 120$\pm$2 & 30.0$\pm$0.5 &  1045 & 2.04 & 1.15 \\
02:40 & 300$\pm$4 & 122$\pm$2 & 27.0$\pm$0.5 &  1160 & 2.17 & 1.14 \\
02:45 & 330$\pm$5 & 124$\pm$2 & 24.0$\pm$0.5 &  1324 & 2.35 & 1.13 \\
02:50 & 380$\pm$6 & 126$\pm$2 & 21.0$\pm$0.5 &  1583 & 2.64 & 1.12 \\
02:55 & 425$\pm$7 & 128$\pm$2 & 18.0$\pm$0.5 &  1841 & 2.92 & 1.11 \\
03:00 & 480$\pm$7 & 130$\pm$2 & 18.0$\pm$0.5 &  2165 & 3.26 & 1.10 \\
03:05 & 545$\pm$8 & 130$\pm$2 & 18.0$\pm$0.5 &  2458 & 3.57 & 1.10 \\
03:10 & 605$\pm$10 & 130$\pm$2 & 18.0$\pm$0.5 &  2729 & 3.86 & 1.10 \\
\hline
03:16 & 1200$\pm$25 & 130$\pm$2 & 18.0$\pm$0.5 & 5413 & 6.69 & 1.10 \\
03:26 & 1400$\pm$36 & 130$\pm$2 & 18.0$\pm$0.5 & 6315 & 7.65 & 1.10 \\
03:37 & 1600$\pm$41 & 130$\pm$2 & 18.0$\pm$0.5 & 7217 & 8.60 & 1.10 \\
03:48 & 1825$\pm$64 & 130$\pm$2 & 18.0$\pm$0.5 & 8232 & 9.68 & 1.10 \\
04:00 & 2100$\pm$83 & 130$\pm$2 & 18.0$\pm$0.5 & 9472 & 11.00 & 1.10 \\
\hline
\end{tabular}
\end{table}

\begin{figure}
   \centering
   \includegraphics[width=8.5cm]{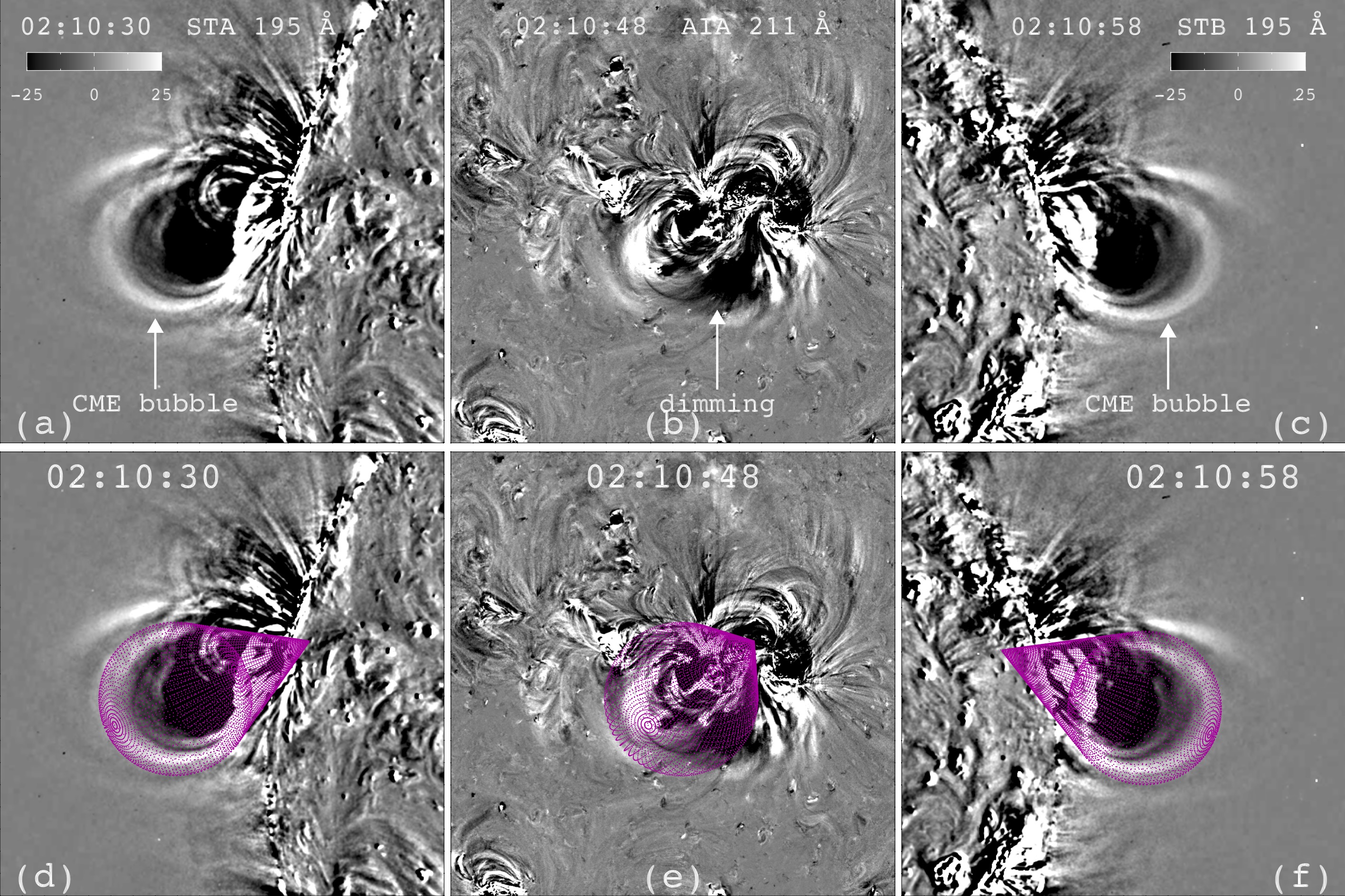}
      \caption{Top panels: base-difference images in 195 {\AA} observed by STA/EUVI (a) and STB/EUVI (c) at 02:10 UT. 
                    Base-difference image in 211 {\AA} observed by SDO/AIA at 02:10 UT (b). The arrows point to the CME bubble and dimming.
                    Bottom panels: the same images as the top panels, with the projections of reconstructed cone being superposed (magenta dots).}
         \label{fig5}
   \end{figure}

In Fig.~\ref{fig5}, base-difference images in 195 {\AA} observed by STA and STB at 02:10 UT are displayed in panels (a) and (c), respectively.
The CME bubble with enhanced intensity is pointed by arrows. 
Base-difference image in 211 {\AA} observed by AIA at 02:10 UT is displayed in panel (b), where projected CME leading edge is blurred and incoherent.
Therefore, the reconstruction of cone is performed using the simultaneous base-difference images of STEREO.
In the bottom panels, the projections of reconstructed cone are superposed with magenta dots. 
It is obvious that the slightly modified cone (Fig.~\ref{fig1}(b)) can nicely fit the CME bubble (see panels (d) and (f)).
The parameters of $r$, $\omega$, $\theta_1$, and the corresponding $l$ (Eqn.~\ref{eqn-5}) are listed in Table~\ref{tab-1}.
Note that the deviation angle $\phi_1=-45^{\circ}$ is derived from the data-constrained MHD simulation \citep{guo21} and keeps constant for convenience.
The initial eruption of flux rope is inclining southward from the radial direction by $\sim$51$^{\circ}$ and the length of CME bubble is $\sim$555$\arcsec$.
To estimate the errors in the derived best-fit cone for a sample snapshot, each parameter is changed for a few times while keeping all the others equal to their best-fit values.
The tolerated offsets from the best-fit cone parameters are considered as uncertainties.

\begin{figure}
   \centering
   \includegraphics[width=8.5cm]{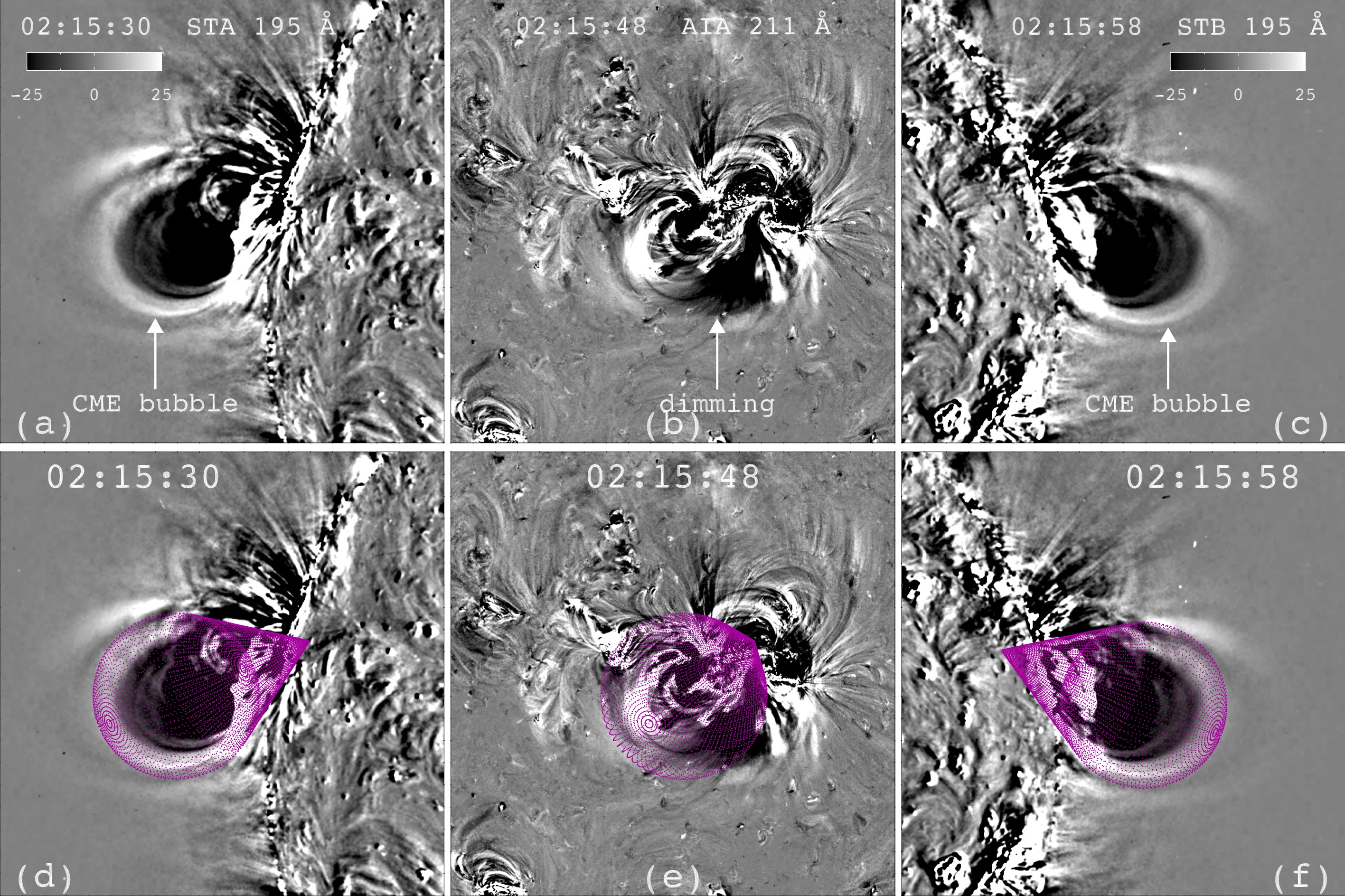}
      \caption{Same as Fig.~\ref{fig5}, but for images at 02:15 UT.}
         \label{fig6}
   \end{figure}

\begin{figure}
   \centering
   \includegraphics[width=8.5cm]{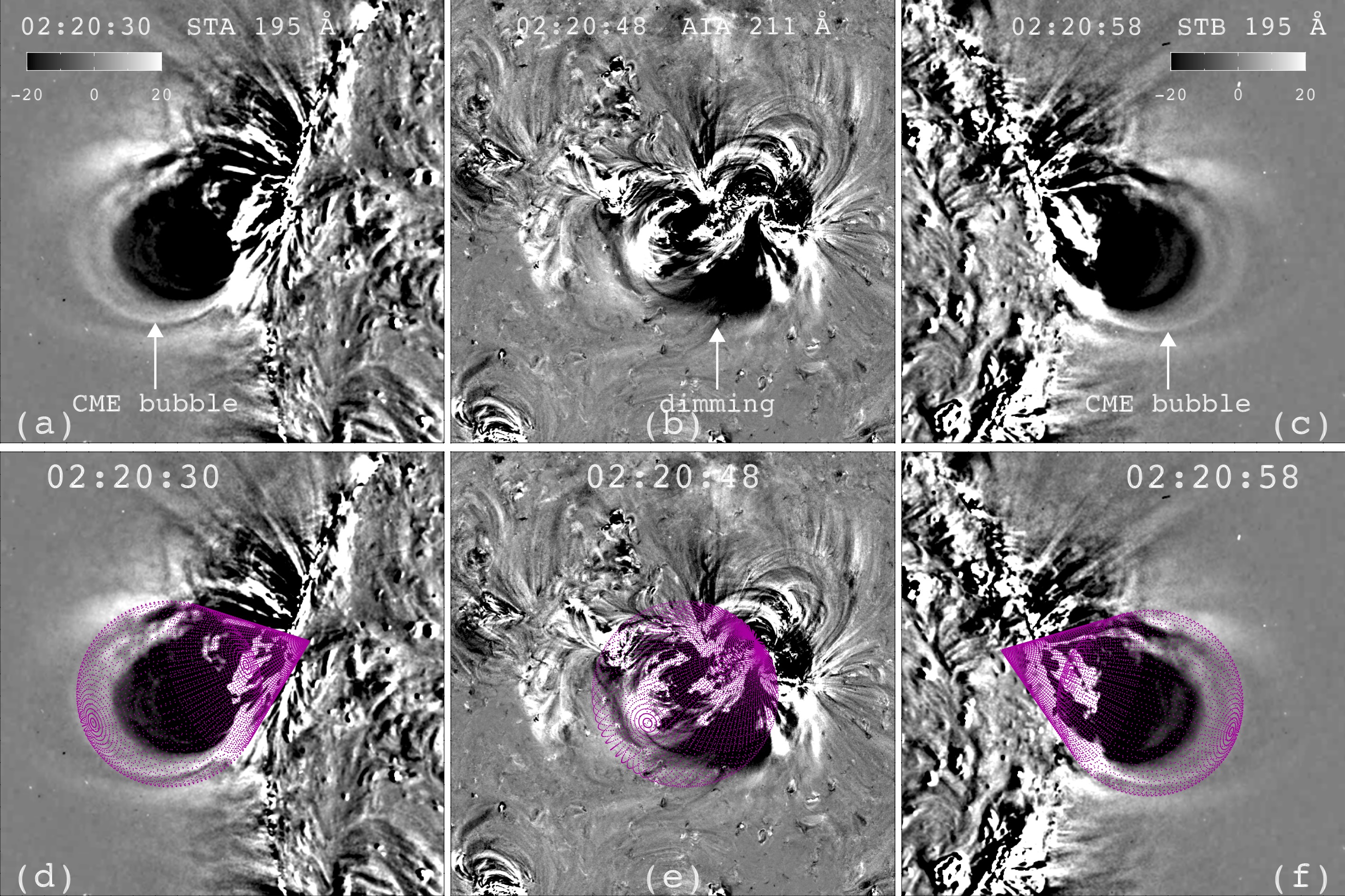}
      \caption{Same as Fig.~\ref{fig5}, but for images at 02:20 UT.}
         \label{fig7}
   \end{figure}

\begin{figure}
   \centering
   \includegraphics[width=8.5cm]{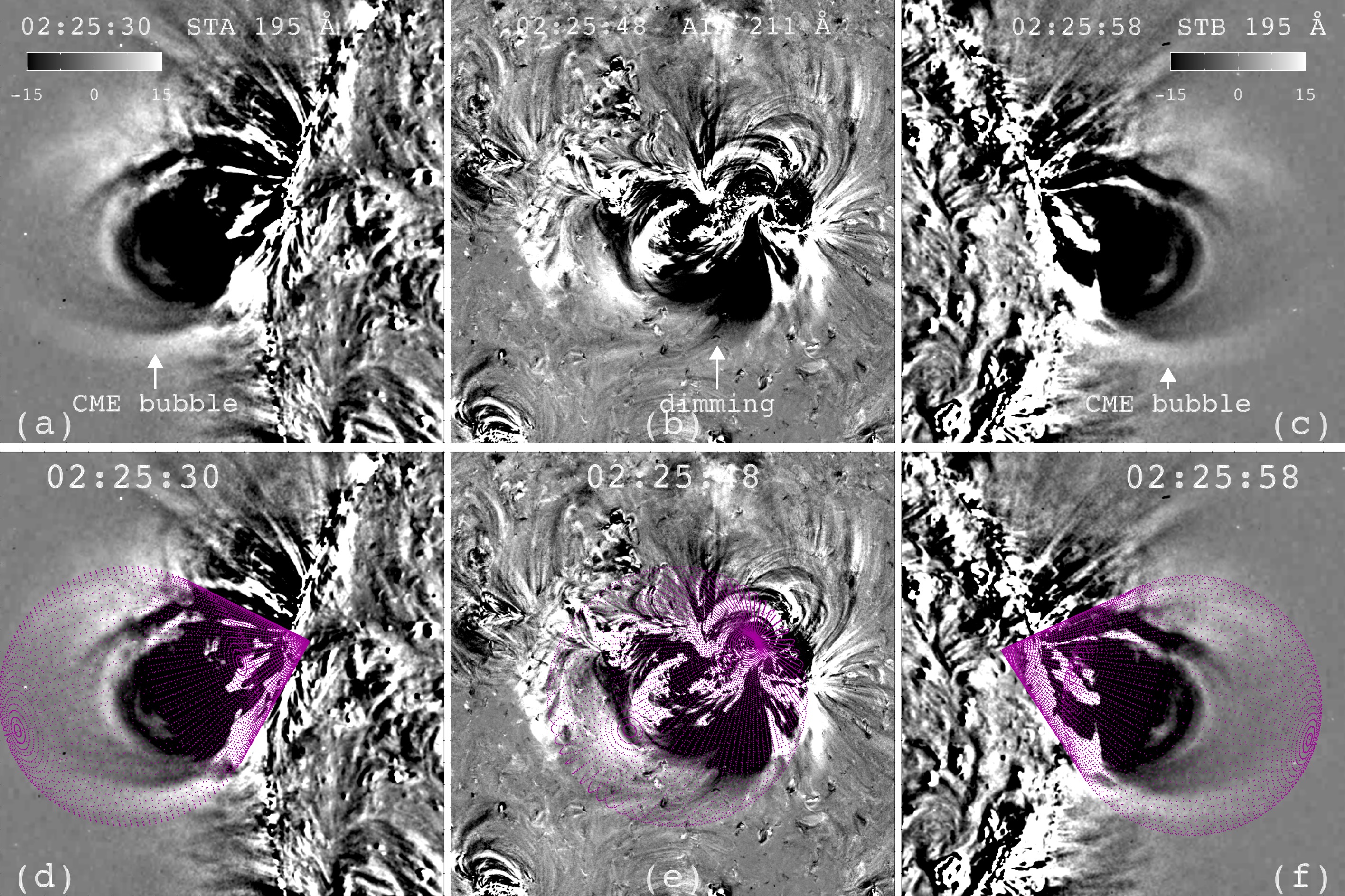}
      \caption{Same as Fig.~\ref{fig5}, but for images at 02:25 UT.}
         \label{fig8}
   \end{figure}

The EUV difference images and projected cones at 02:15 UT, 02:20 UT, and 02:25 UT are demonstrated in Fig.~\ref{fig6}, Fig.~\ref{fig7}, and Fig.~\ref{fig8}, respectively.
It is clear that the cone in Fig.~\ref{fig1}(b) can still fit the CME bubbles within the EUVI FOV. In the AIA FOV, the projected cones cover most of the corresponding dimming regions on the disk.
The derived parameters are listed in Table~\ref{tab-1}. 
The cone angle ($\omega$) increases from $\sim$54$^{\circ}$ to $\sim$76$^{\circ}$ within 15 minutes at a rate of $\sim$1.5$\degr$ minute$^{-1}$, indicating a rapidly lateral expansion.
The inclination angle ($\theta_1$) decreases from $\sim$51$^{\circ}$ to $\sim$44$^{\circ}$ at the same time, suggesting a tendency of radial propagation.
The length of leading edge ($l$) increases gradually from 555$\arcsec$ to 769$\arcsec$. 
The heliocentric distance ($d$) of the leading edge increases from $\sim$1.44\,$R_\odot$ to $\sim$1.68\,$R_\odot$ accordingly (see Fig.~\ref{fig2}(b-d)).
It is noted that CMEs and EUV waves are frequently related \citep{chen11}. 

\begin{figure}
   \centering
   \includegraphics[width=8.5cm]{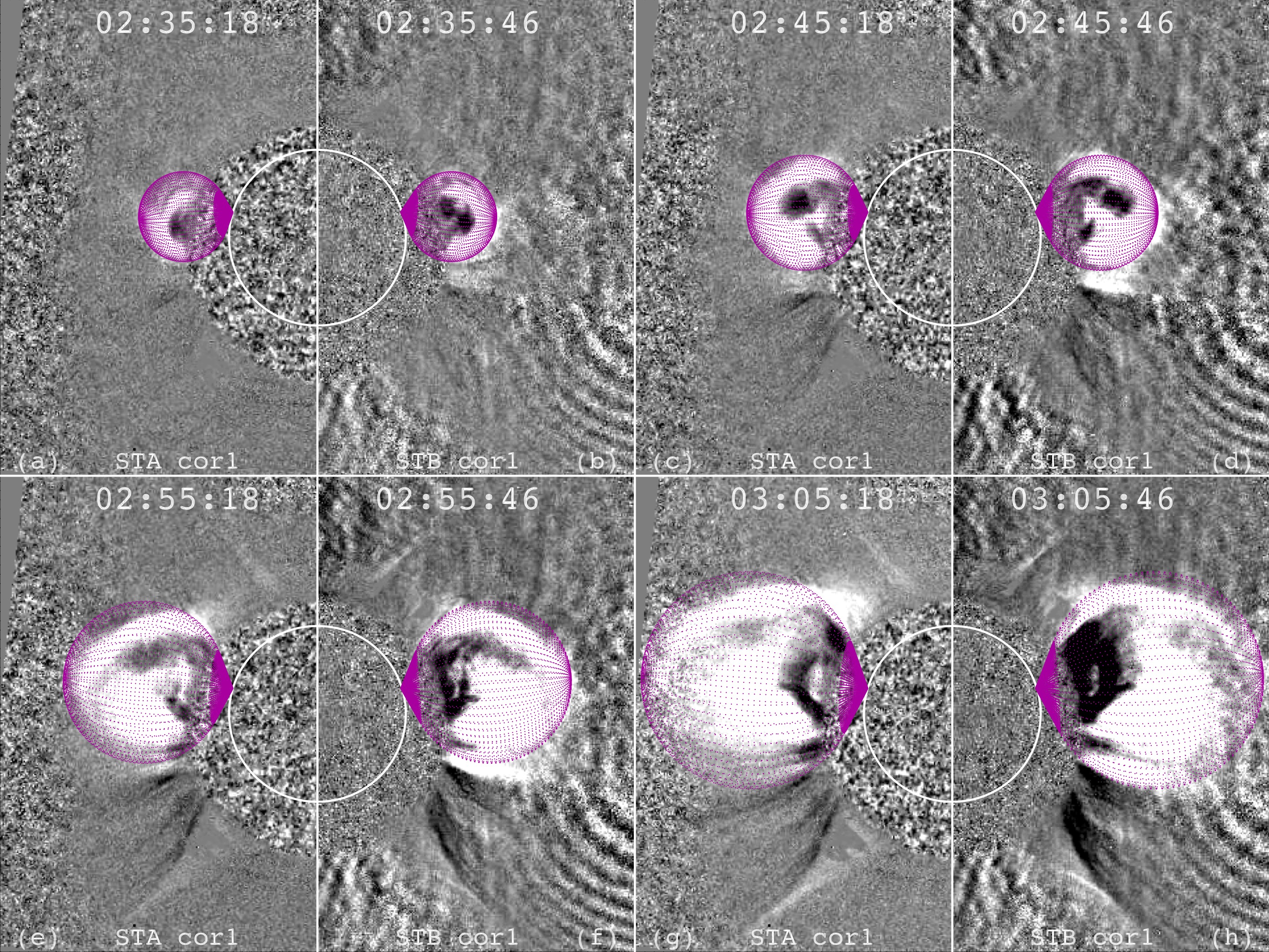}
      \caption{Selected base-difference images of the CME observed by STEREO/COR1 during 02:30$-$03:10 UT.
                    The projections of reconstructed cones are superposed with magenta dots.}
         \label{fig9}
   \end{figure}

During 02:30$-$03:10 UT, the CME leading edge has escaped the STEREO/EUVI FOV and entered the STEREO/COR1 FOV. 
However, the CME does not appear in the LASCO/C2 FOV until 03:16 UT.
Figure~\ref{fig9} shows selected base-difference images of the CME observed by COR1 during 02:30$-$03:10 UT. The projections of reconstructed cones are superposed with magenta dots.
The parameters of cone are listed in Table~\ref{tab-1}. The cone angle increases gradually from $\sim$110$^{\circ}$ to $\sim$130$^{\circ}$ within half an hour and reaches a plateau.
The inclination angle decreases from $\sim$36$^{\circ}$ to $\sim$18$^{\circ}$ in the meanwhile.
The total length of the leading edge increase quickly from $\sim$793$\arcsec$ to $\sim$2729$\arcsec$, 
and the corresponding heliocentric distance increases from $\sim$1.75\,$R_\odot$ to $\sim$3.86\,$R_\odot$ (see Fig.~\ref{fig2}(b-d)).

\begin{figure}
   \centering
   \includegraphics[width=8.5cm]{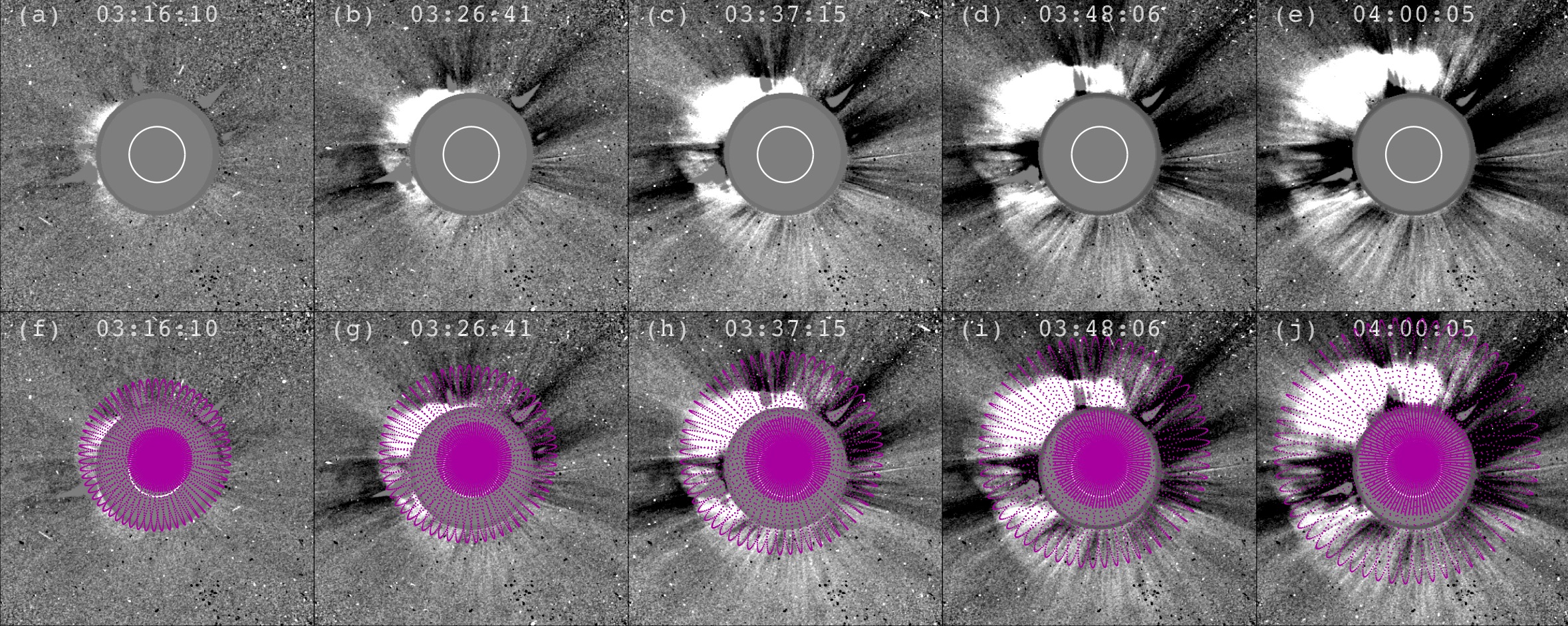}
      \caption{Top panels: base-difference images of the CME observed by LASCO/C2 during 03:16$-$04:00 UT.
                    Bottom panels: the same difference images with projections of reconstructed cones being superposed with magenta dots.}
         \label{fig10}
   \end{figure}

After 03:15 UT, the CME leading edge escapes the COR1 FOV and enters the C2 FOV. Consequently, there is no constraint of the cone from STEREO observation any more. 
Assuming that the CME shape remains self-similar and the direction keeps constant in the LASCO FOV, 
the cone angle is set to be 130$^{\circ}$ and the inclination angle is set to be 18$^{\circ}$ (see Table~\ref{tab-1}).
In Figure~\ref{fig10}, the top panels show base-difference images of the CME observed by LASCO/C2 during 03:16$-$04:00 UT. 
In the bottom panels, the projections of reconstructed cones are superposed with magenta dots, which could moderately cover most of the CME.
The length of the cone increases from $\sim$5400$\arcsec$ to $\sim$9500$\arcsec$, 
and the corresponding heliocentric distance increases from $\sim$6.7\,$R_\odot$ to $\sim$11.0\,$R_\odot$ (see Fig.~\ref{fig2}(b-d)).
The real speed of the CME leading edge is calculated to be $\sim$1139 km s$^{-1}$. 
In Fig.~\ref{fig2}(b), the heights of CME in LASCO/C2 FOV are drawn with chocolate triangles, with an apparent linear speed of $\sim$724 km s$^{-1}$.
Hence, the ratio of real speed to the apparent speed of CME is $\sim$1.6. Such a high speed of CME is likely to drive a shock wave \citep{on09}. 
In Fig.~\ref{fig10}, the CME leading front might be blended with the shock front, which makes it difficult to distinguish the CME front.
In Fig.~\ref{fig2}(b), the leap of heliocentric distance at 03:16 UT is probably caused by an overestimation of CME height when constraint from another viewpoint is unavailable.
However, a type II radio burst accompanying the shock wave is absent in the radio dynamic spectra during the eruption. An in-depth investigation is needed to address this issue \citep{rou12}.
   
\begin{figure}
   \centering
   \includegraphics[width=8.5cm]{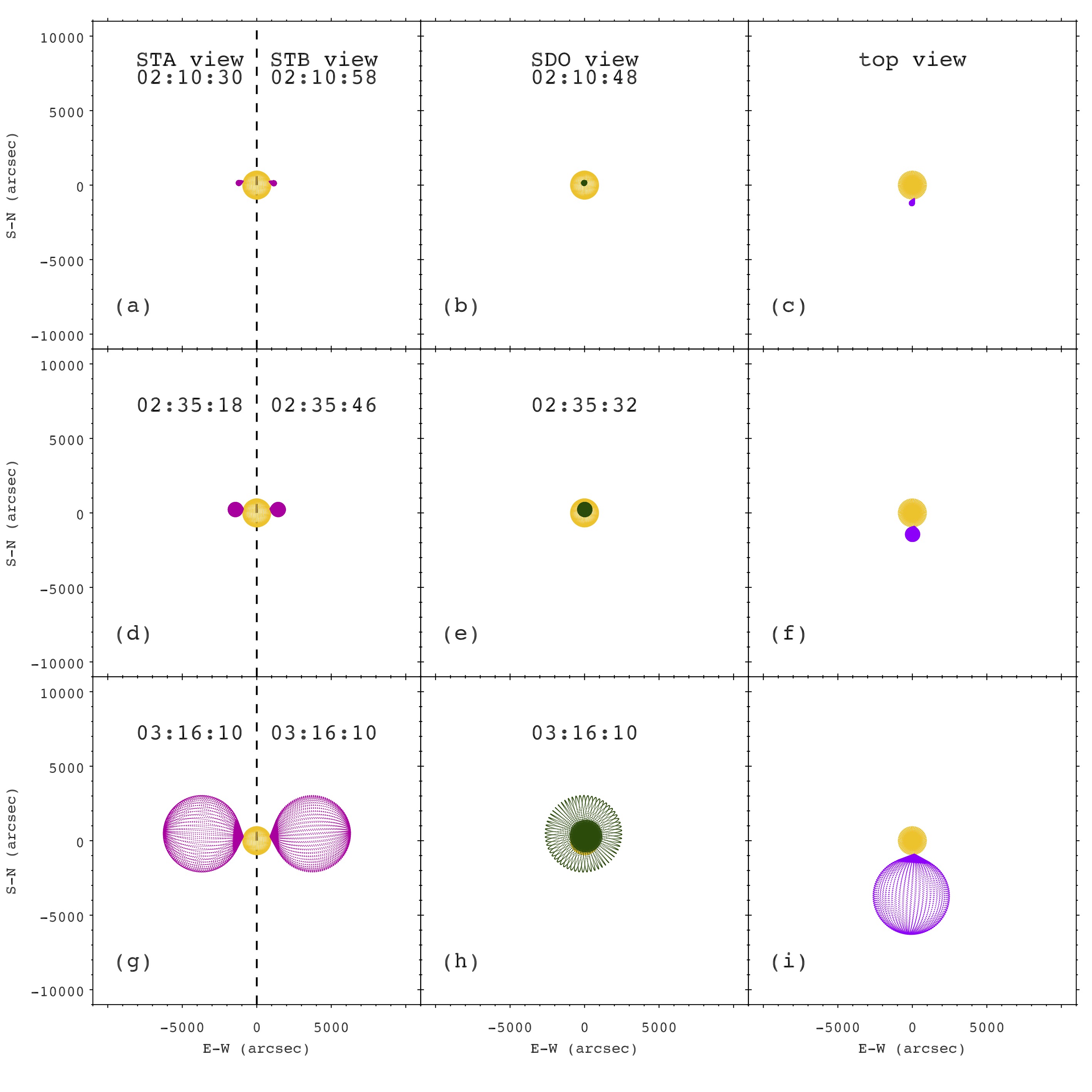}
      \caption{Snapshots of the Sun (yellow dots) and reconstructed CME cones from viewpoints of STA (magenta dots), STB (magenta dots), SDO (green dots), and north pole (purple dots).
                   An animation (\textsf{cones.mp4}) is available online.}
         \label{fig11}
   \end{figure}

Figure~\ref{fig11} shows snapshots of the Sun (yellow dots) and reconstructed CME cones from viewpoints of STA (magenta dots), STB (magenta dots), SDO (green dots), 
and north pole (purple dots), respectively. The whole evolution during 02:10$-$04:00 UT is demonstrated in an online animation (\textsf{cones.mp4}).
The non-radial eruption of the prominence-carrying flux rope is like blowing a bubble. The time variations of $\omega$, $\theta_1$, and $d$ are plotted in Fig.~\ref{fig2}. 
The cone angle increases sharply during 02:10$-$02:40 UT, which is coincident with the fast rise of the flux rope (Fig.~\ref{fig2}(a)), indicating a rapid expansion and propagation in the initial phase.
\citet{pat10b} studied the early evolution of the eruptive flare on 2010 June 13. A short-lived, lateral overexpansion with a declining aspect ratio is discovered in the deceleration phase.
\citet{ver18} investigated the genesis of an extremely fast CME on 2017 September 10, 
finding that the strong lateral overexpansion of CME bubble drives the fast-mode EUV (shock) wave during the impulsive phase of the related X8.2 class flare.
In the current case, the aspect ratio declines from $\sim$2.2 at 02:10 UT to $\sim$1.1 at 02:55 UT, which is accordant with the lateral expansion of CME bubble (Fig.~\ref{fig2}(c)).
The heliocentric distance of CME leading front increases gradually in the initial phase and quickly in the later phase up to $\sim$11\,$R_{\odot}$.

\section{Discussion} \label{dis}
Since CMEs are the most spectacular driver of space weather on Earth, the morphology, propagation, and kinematics of CMEs are crucial to an accurate prediction of space weather.
The 3D reconstructions of CMEs using multi-point observations are substantial \citep[e.g.,][]{mor04,by10,feng12,rou16}. \citet{by10} used an elliptical tie-pointing technique to 
reconstruct a full CME front in 3D. The angular width increases with the heliocentric distance from 2\,$R_{\odot}$ to 46\,$R_{\odot}$ with a power law form. 
The deflection from the radial from the onset decreases with distance as well, suggesting an equator-ward deflection. 
\citet{shen11} investigated the kinematic evolution of the CME on 2007 October 8. The CME first deflected to the lower latitude by $\sim$30$^{\circ}$ and then propagated radially.
\citet{gui11} analyzed the deflections of ten CMEs observed by the STEREO twin spacecraft and concluded that the deflections are primarily controlled by the background magnetic field.
\citet{is14} explored the whole propagations of 14 CMEs from the Sun to 1 AU, finding that the deflection of flux ropes occurs below 30\,$R_{\odot}$ in most cases.
\citet{zqm21} applied the revised cone model to two CMEs as a result of non-radial prominence eruptions. Similarly, the inclination angles are 60$^{\circ}-$70$^{\circ}$ at the very beginning, 
and the directions of CMEs become radial in the LASCO FOV. In the current event, the halo CME experiences lateral expansion during its eruption.
A crude fitting between $\omega$ and $d$ results in a relationship of $\omega=130\degr-480d^{-5}$, where $d$ is in unit of $R_{\odot}$.
The inclination angle ($\theta_1$) decreases from $\sim$51$^{\circ}$ to $\sim$18$^{\circ}$, suggesting a trend of radial propagation.
A crude fitting between $\theta_1$ and $d$ results in a relationship of $\theta_1=81.7\degr-23.45d$, where 1.4\,$R_{\odot}$ $\leq d \leq3.0$\,$R_{\odot}$.

Compared with the preliminary application in \citet{zqm21}, the reconstructions of CME extend to the LASCO/C2 FOV, which is a step forward.
It should be emphasized that the performance still has limitations. Firstly, the deviation angle ($\phi_1$) is assumed to keep constant for simplicity.
Besides, the angular width and direction of CME are assumed to keep constant in the LASCO/C2 FOV, since the leading edge has escaped the COR1 FOV.
The not exactly satisfactory fitting of the C2 data might not only arise from the possible confusion with the shock wave, but due to the fact that the model is applied to a single viewpoint as well.
Secondly, the time cadence of LASCO/C2 is $\geq$10 minutes, which is twice lower than EUVI and COR1.
Finally, considering the dispersed shape and weak intensity of CME in the LASCO/C3 FOV, the reconstruction is not performed after 04:00 UT.

In the future, a comprehensive tracking of the CME propagation is feasible in combination with the GCS modeling \citep{the06}.
The revised cone model will be applied to 3D reconstruction of CMEs observed by the Lyman-$\alpha$ Solar Telescope \citep[LST;][]{feng19,li19} on board 
the Advanced Space-based Solar Observatory \citep[ASO-S;][]{gan19}, the Metis \citep{ant20} and Heliospheric Imager \citep[SoloHI;][]{how20} on board Solar Orbiter \citep{mu20}.
The model is also valuable in studying fast CMEs driving shock waves, with additional constraints from radio observations \citep[e.g.,][]{go09,zu18,man19}.
The reconstructed morphology and direction of a CME may serve as the initial condition for MHD numerical simulations \citep{shen21,yang21}.

\section{Summary} \label{sum}
In this paper, the revised cone model is slightly modified and applied to a full halo CME as a result of non-radial flux rope eruption on 2011 June 21.
The main results are as follows:
   \begin{enumerate}
      \item The cone shape fits well with the CME observed by EUVI and COR1 on board STEREO twin spacecraft and LASCO/C2 coronagraph.
      \item The cone angle increases sharply from $\sim$54$^{\circ}$ to $\sim$130$^{\circ}$ in the initial phase, indicating a rapid expansion.
      A relation between the cone angle and heliocentric distance of CME leading front is derived, $\omega=130\degr-480d^{-5}$, where $d$ is in unit of $R_{\odot}$.
      The inclination angle decreases gradually from $\sim$51$^{\circ}$ to $\sim$18$^{\circ}$, suggesting a trend of radial propagation.
      The heliocentric distance increases gradually in the initial phase and quickly in the later phase up to $\sim$11\,$R_{\odot}$.
      The true speed of CME reaches $\sim$1140 km s$^{-1}$, which is $\sim$1.6 times higher than the apparent speed in the LASCO/C2 FOV.
      \item The revised model is promising in tracking the complete evolution of CMEs.
   \end{enumerate}

\begin{acknowledgements}
The author appreciates the reviewer for valuable suggestions to improve the quality of this article.
The author is also grateful to Drs. Y. Guo and Z. J. Zhou for helpful discussions.
SDO is a mission of NASA\rq{}s Living With a Star Program. AIA data are courtesy of the NASA/SDO science teams.
STEREO/SECCHI data are provided by a consortium of US, UK, Germany, Belgium, and France.
This work is funded by the National Key R\&D Program of China 2021YFA1600500 (2021YFA1600502), 
NSFC grants (No. 11773079, 11790302), and the Strategic Priority Research Program on Space Science, CAS (XDA15052200, XDA15320301).
\end{acknowledgements}

\end{document}